\documentclass[useAMS,usenatbib]{mnras}

\usepackage{geometry}                		
\usepackage{graphicx}				
\usepackage{amssymb}

\usepackage{amsmath}
\usepackage{epsfig}
\usepackage{hyperref}

\newcommand{\kms}{km\,s$^{-1}$ }
\newcommand{\kmsns}{km\,s$^{-1}$}
\newcommand{\kmsikpc}{km\,s$^{-1}$\,kpc$^{-1}$ }
\newcommand{\kmsikpcns}{km\,s$^{-1}$\,kpc$^{-1}$}

\title[]{Spiral arm crossings inferred from ridges in {\it Gaia} stellar velocity distributions}
\author[Quillen et al.]{
Alice C. Quillen$^{1,2}$,
Ismael Carrillo$^2$,
Friedrich Anders$^2$,
Paul McMillan$^3$,
\newauthor
Tariq Hilmi$^{2,4}$, 
Giacomo Monari$^2$,
Ivan Minchev$^2$,
\newauthor
Cristina Chiappini$^2$,
Arman Khalatyan$^2$,
and
Matthias Steinmetz$^2$
\\
$^1$Department of Physics and Astronomy, University of Rochester, Rochester, NY 14627, USA\\
$^2$Leibniz Institut f\"ur Astrophysik Potsdam (AIP), An der Sternwarte 16, D-14482, Potsdam, Germany\\
$^3$Lund Observatory, Lund University, Department of Astronomy and Theoretical Physics, Box 43, SE-22100, Lund, Sweden\\
$^4$Astrophysics Research Group, University of Surrey, Guildford, Surrey GU2 7XH, UK\\
}

\hypersetup{draft}

\begin{document}
\maketitle

\begin{abstract}
The solar neighborhood contains disc stars that have recently crossed spiral arms in the Galaxy. We propose that  boundaries  in local velocity  distributions separate stars that have recently crossed and been more strongly perturbed by  a particular arm from those that haven't.  Ridges in the stellar velocity distributions constructed from the second {\it Gaia} data release trace orbits that could have touched nearby spiral arms at apocentre or pericentre. The multiple ridges and arcs seen in local velocity distributions are consistent with the presence of multiple spiral features and different pattern speeds and imply that the outer Galaxy is  flocculent rather than grand design.

\end{abstract}

keywords: galaxies: kinematics and dynamics

\section{Introduction}

Stars in Milky Way  disc can be described in terms of
their distributions in space and velocity or phase space. 
Perturbations on the disc can cause structures such as gaps and arcs within this space. 
Non-axisymmetric perturbations from spiral arms \citep{quillen05} or the Galactic bar 
can induce a gap in a local velocity distribution associated with an orbital
 resonance \citep{kalnajs91,dehnen99b,fux01,minchev07,minchev10,
monari14,antoja14,antoja15,monari17a,monari17b,perez17,quillen18,hunt18}.  
Dissection of an N-body simulation suggests that
gaps in velocity distribution might be seen all over the Galaxy and 
are associated with transitions between different spiral patterns \citep{quillen11}.  
\citet{hunt17}  proposed that high angular momentum stars 
 are affected by the Perseus spiral arm which lies 2 kpc outside the Sun's galactocentric radius
 (e.g., \citealt{xu16}).
Perturbations on the outer disc, such as by the Sagittarius dwarf galaxy,
can cause correlated radial epicyclic oscillations giving arcs
in phase space \citep{minchev09}, clumps in velocity distributions \citep{quillen09},
and vertical motions that can be correlated with planar ones
\citep{gomez13,delavega15,quillen18,monari18,antoja18,laporte18}.

The second data release (DR2) \citep{gaiaDR2} of the {\it Gaia} satellite \citep{gaia} 
provides full 6-dimensional space coordinates for approximately 7 million stars in the Galaxy:
2D positions (RA, Dec on the sky), 
parallaxes, proper motions 
\citep{lindegren18},
and radial line-of-sight velocities for magnitude $G_{\rm RVS}<12$  \citep{katz18}. 
With the increase in numbers of stars and precision of measurement, {\it Gaia} DR2 lets us examine
stellar velocity distributions, and how they vary with position in the Galaxy,  in unprecedented detail.
We attempt to better  characterize the dynamical processes causing structure in the 
stellar velocity distributions.

\subsection{Interpreting arcs and shells in velocity distributions}

To discuss velocity distributions,
we use Galactocentric polar coordinates $(r,\theta,z)$ giving velocity components $(v_r, v_\theta, v_z)$.
Nearby star motions are often described with a Cartesian coordinate
system with velocity components $u,v,w$.  At the position of the Sun,  $u=-v_r$
$v=v_\theta - v_c$, and $w = v_z$,  
with $v_c$ the rotational velocity associated with the local standard of rest (LSR).

Arcs and shells are visible in the {\it Gaia} DR2 Galactic disc's $v_r, v_\theta$ velocity distribution, 
and ridges are present in space-velocity diagrams \citep{katz18,antoja18,kawata18}.
These features could be caused by spiral arms  \citep{kawata18}
or phase wrapping of epicyclic motions 
associated with past disturbances of the Milky Way disc  \citep{antoja18}.
The substructure seen in phase space implies that the disc is not dynamically relaxed
\citep{minchev09,gomez12,monari18,katz18,antoja18}.   
Using the sensitivity of the Coma Berenices moving group to viewed Galactic hemisphere \citep{quillen18}, 
\citet{monari18} estimated
 that the disc experienced a vertical perturbation 1.5 Gyr ago.
\citet{antoja18} interpreted
a spiral seen in $v_z$ vs $z$ plots of stars in the solar neighborhood as phase wrapping of a 
vertical perturbation that occurred between 300 and 900 Myr ago.  
The phase wrapping model by \citet{minchev09} explained arcs in the $uv$ velocity distribution
by weighting the phase space distribution with a function that depends on the
period of epicyclic oscillations. 
The velocity distribution in the solar neighborhood's has streaks separated by 10 to 20 \kms giving
a timescale since perturbation of  2 to 4 Gyr (following Figure 1 by \citealt{minchev09}).  
The three timescales are not consistent and
imply that additional processes, such as bar or spiral arms, affect motions in the plane.

The arcs predicted with the phase wrapping model 
by \citet{minchev09} depend only on the frequency of radial epicyclic oscillations.
As the period of radial oscillations
is primarily dependent on the orbital energy \citep{dehnen99} ($\propto v_r^2 + v_\theta^2$), the predicted
arcs resemble large circles centred
about $v_r=0, v_\theta = 0$ and so are symmetrical about the $v_r=0$ line.
However the arcs seen in the solar neighborhood  velocity distribution 
are tilted with respect to this line (see Figure 22 by \citealt{katz18}).
Bar and spiral resonant models predict a tilt in the orientation of arcs in the velocity distribution
\citep{quillen05,minchev10,monari17a,monari17b} but only
near a resonance.  
The tilts of so many arcs seen in the {\it Gaia} DR2 solar neighborhood 
 velocity distribution suggest that they cannot  be explained by orbital resonances alone.

Spiral arms in the Galaxy are tightly wound and separated by about 2 kpc with the Perseus
arm currently about 2 kpc outside the galactocentric radius of the Sun, the Sagittarius arm about 2 kpc
inside the radius of the Sun, the Local Arm about 500 pc outside the Sun's radius (sometimes called
the Orion Spur) and the Local Spur (seen at Galactic longitude $\sim 40^\circ$) 
a few hundred pc within it; following maser sources plotted in Figure 2 by \citet{xu16} (but also see \citealt{russeil03,reid09,xu13,houhan14,vallee17}).
In an N-body simulation of a disc exhibiting spiral structure (and without external perturbations), 
multiple arcs that are tilted with respect to the $v_r=0$ line 
are sometimes and in some places seen in  local velocity 
distributions  \citep{quillen11}.
This motivates searching for a spiral arm related explanation for arcs or ridges in the solar neighborhood's
 velocity distribution.

A  star near the Sun  on an eccentric orbit moves away from the solar neighborhood and can cross
one of these nearby
spiral arms.  We consider the possibility that some of the structure
seen in the local velocity distribution could be due to stars crossing nearby spiral arms.
An analogy is the perihelion boundary marking possible close approaches with Neptune that
is present in the distribution of semi-major axes and eccentricities of non-resonant Edgeworth-Kuiper belt objects.\footnote{\url{http://www.scholarpedia.org/article/Kuiper_belt_dynamics}}
A spiral arm might give a boundary
in phase space, where on one side of the boundary stars never cross the arm 
and are only weakly perturbed by it,
and on  the other side, their velocities are more strongly perturbed by the arm.
Bumps and wiggles  in the terminal velocity
curve of the Milky Way  observed interior to the solar circle 
correspond to known spiral arm features \citep{mcgaugh16}.  The bumps 
have sizes of $\pm$ 10--20 \kms \citep{mcgaugh16}.   A star crossing
a spiral arm would experience a jump in velocity about this size,
implying that 
that spiral arms must influence the stellar velocity distribution.
N-body simulations show that perturbations at  apocentre are strongest from
spiral structures with  pattern speed approximately corotating with galactic rotation (e.g., \citealt{kawata14}).

In this manuscript we explore a possible connection between Galactic spiral structure and arcs or boundaries
in the solar neighborhood's velocity distribution.   In section \ref{sec:spiral}
we estimate the time and location of a star's last
pericentre or apocentre, relating recent orbital extrema to velocity vectors.  For logarithmic spirals,
we estimate the current location of a spiral arm that would have been touched at apocentre
and pericentre.   In section \ref{sec:multiple} we associate ridges in the solar neighborhood's
velocity distribution to possible spiral arms, and test how the locations of
 features in the velocity distribution depends on neighborhood position near the Sun.
A summary and discussion follows in section \ref{sec:sum}.
 
\begin{figure}
\includegraphics[width=8.0cm,trim={20mm 0 0 0},clip]{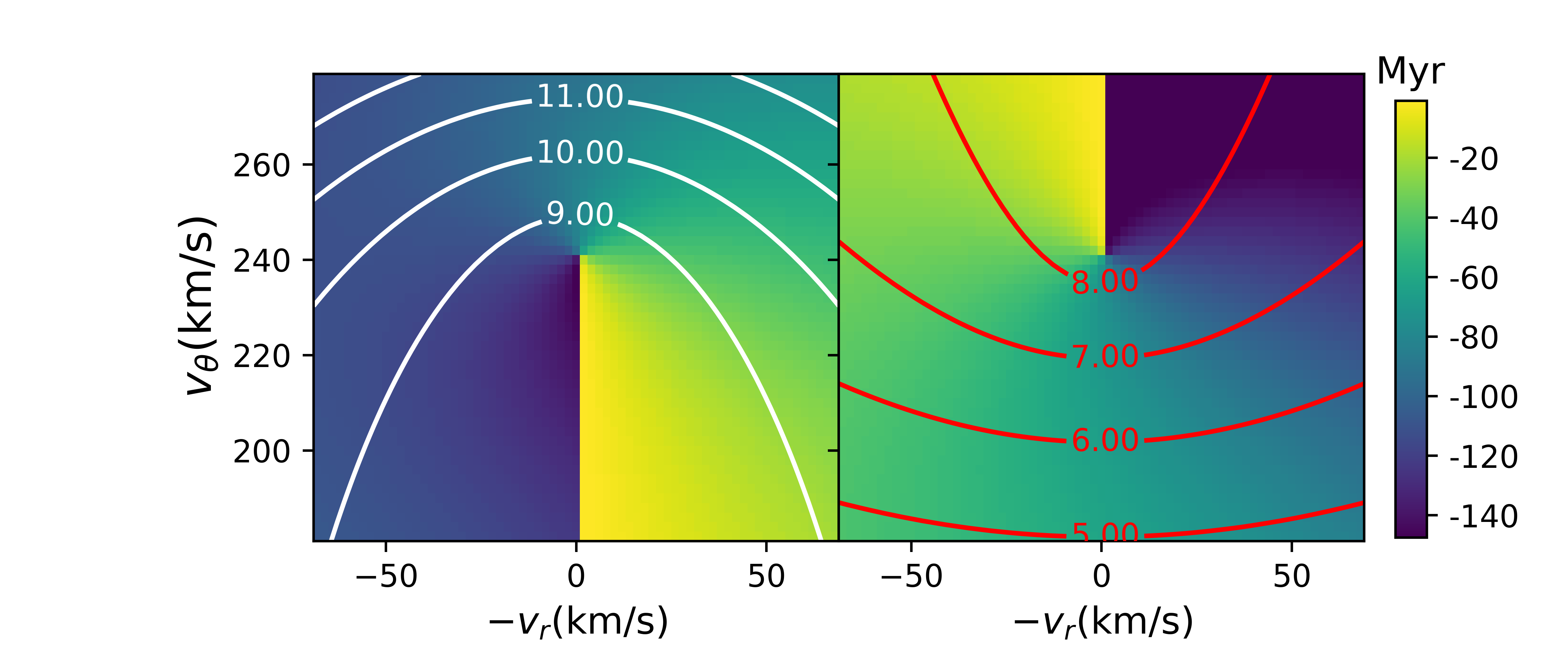}
\caption{Past apocentre and pericentre radii and times for stars with different tangential and radial velocities
in the solar neighborhood.  The left panel shows time since the previous apocentre in color
and with apocentre radii shown as contours.  The right panel shows time since the previous
pericentre in color with pericentre radii shown as contours.
These are computed by integrating the orbits in the Galactic plane 
backwards from the location of the Sun and assuming a flat rotation curve.
\label{fig:ecc}}
\end{figure}

\begin{figure}
\includegraphics[width=8.0cm,trim={0 0 0 0},clip]{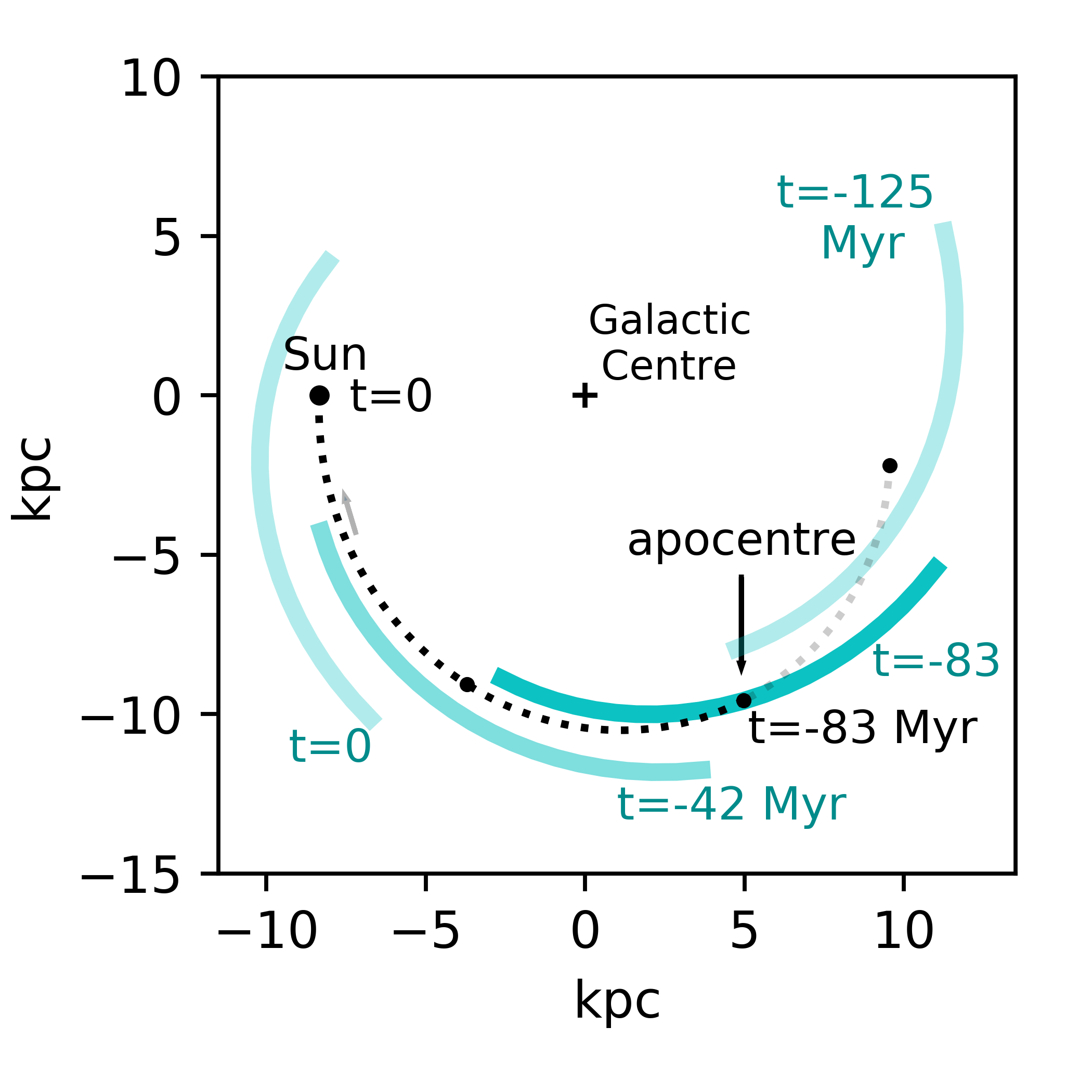}
\caption{A star currently near the Sun and with velocity components 
$v_r=-10$ \kms and $v_\theta = 271$ \kmsns,
reached apocentre approximately 83 Myr ago at the location of the black dot on the lower right.
The star's orbit is shown with the dotted black line and was integrated assuming a flat rotation curve.
A spiral arm like the Perseus arm  currently lies outside the radius of the Sun,
however 83 Myr ago, this arm could have intersected the star's orbit when the star reached apocentre.
The spiral arm has $\alpha =5$, a pattern speed of $\Omega_s = 20$ \kmsikpcns, and
a radius of $R_{s0} = 10$ kpc at the current time and at $\theta = \theta_\odot$.
It is shown as thick cyan lines at the current time, 42,   83 and 125 Myr ago. 
 Black dots are also shown for the star at the same times.  The star remains
inside the arm,  grazing it only at apocentre. 
\label{fig:illustration}}
\end{figure}

\begin{figure*}
\includegraphics[width=11.25cm,trim={0mm 2mm 0mm 2mm},clip]{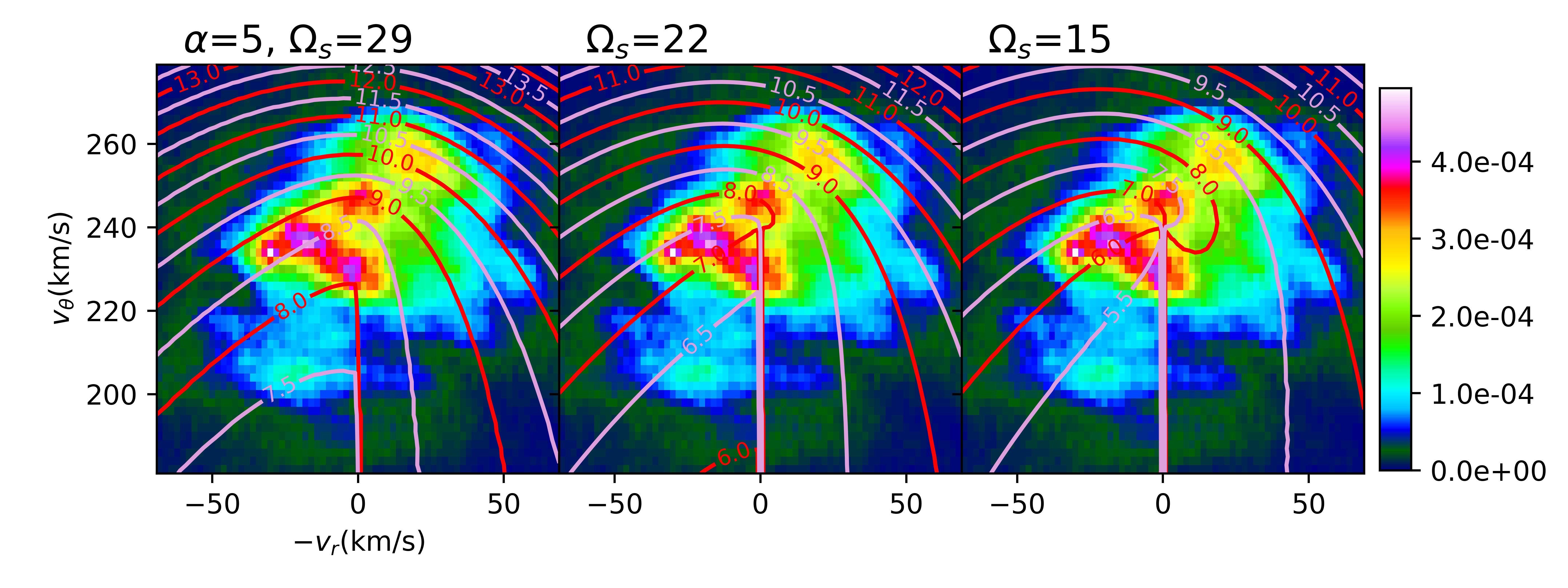}
\includegraphics[width=11.25cm,trim={0mm 2mm 0mm 2mm},clip]{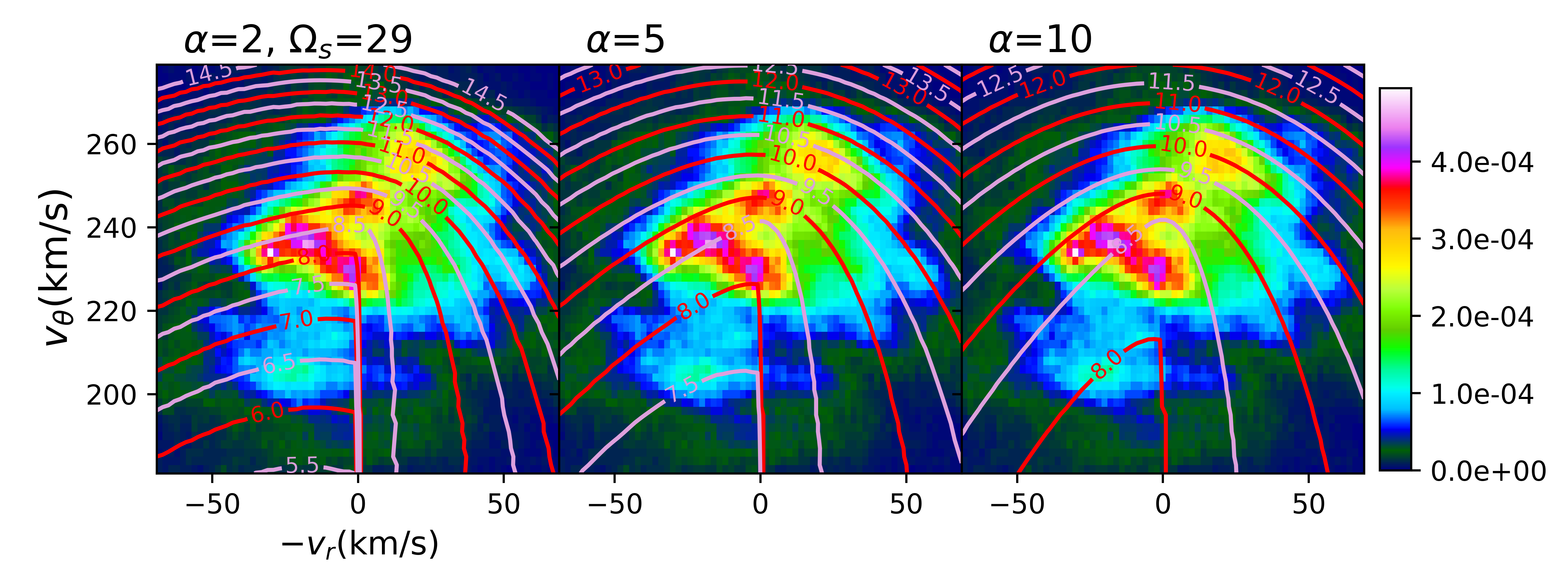}
\includegraphics[width=11.25cm,trim={0mm 2mm 0mm 2mm},clip]{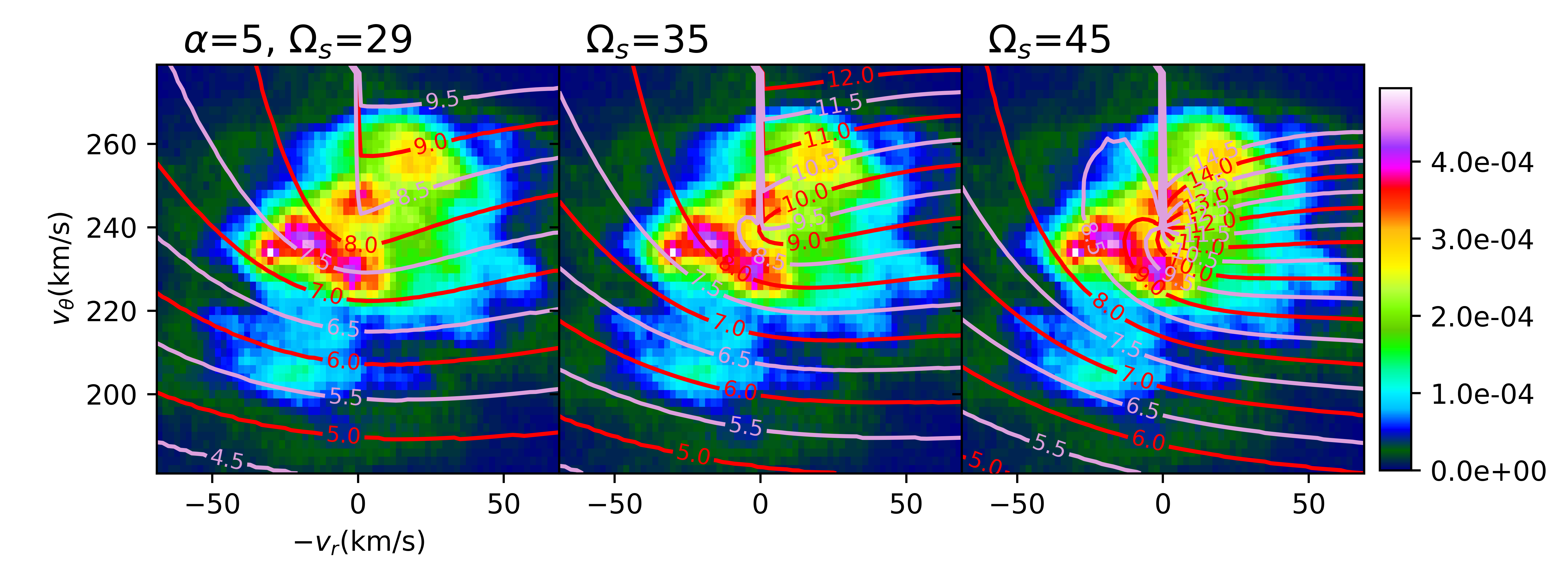}
\includegraphics[width=11.25cm,trim={0mm 2mm 0mm 2mm},clip]{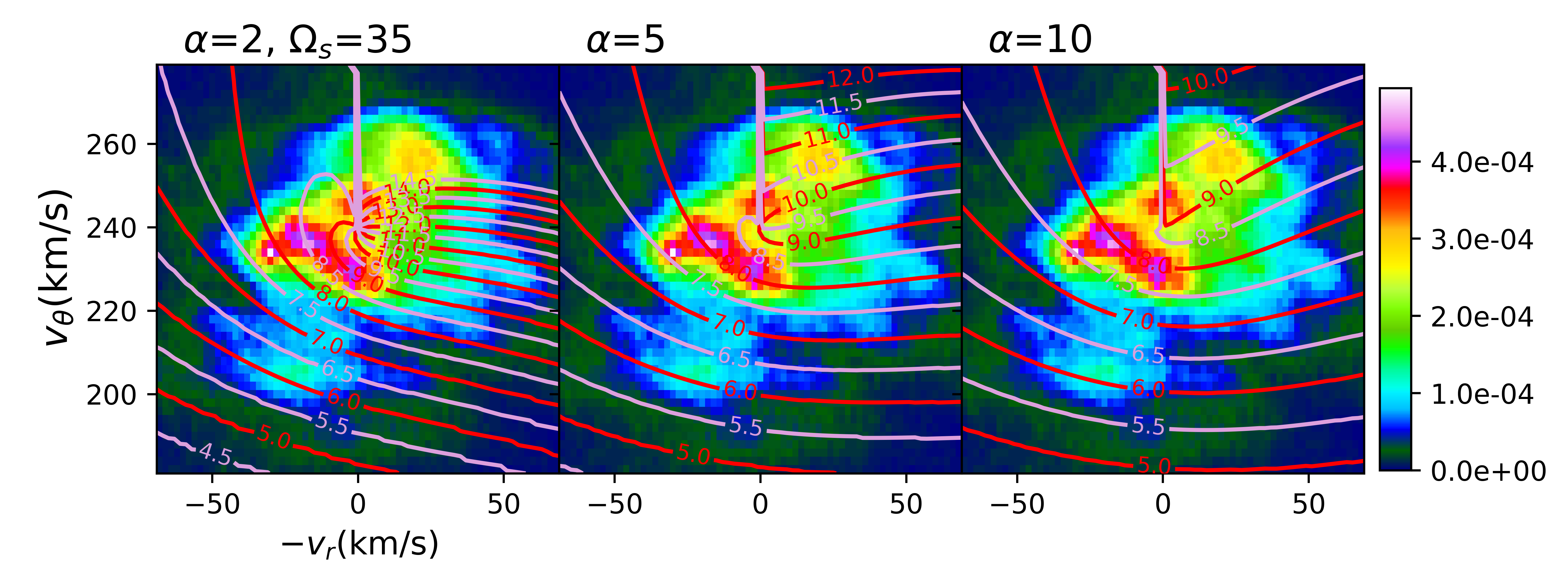}
\caption{Top row:  Colored Contours are curves of constant spiral radius $R_{s0}$ for spiral arms that were reached at apocentre
for stars currently at the position of the Sun and as a function of velocity vector.  The $x$-axis shows
the radial velocity component and the $y$-axis the tangential velocity component (in \kmsns) of the star's present
velocity.   The contours  are 
 labelled in kpc.  The underlying color images
show the normalized stellar velocity distribution in the solar neighborhood constructed 
using a histogram of stars with measurements
from {\it Gaia} DR2.  The number density of stars is shown with the color map
on the right.   The winding parameter for the spiral pattern is $\alpha  =5$.   Pattern speeds
for the spiral arms are given on the top of each panel in units of \kmsikpcns.
Second from top row:   Similar to the top row, but the winding parameter is varied instead of the pattern speed.
Winding parameters for the spiral arms are labelled on top of each panel.
Third from top row:  Similar to the top row but for spiral patterns that are reached at pericentre.
Bottom row: Similar to the second from top row but for spiral patterns that are reached at pericentre.
\label{fig:maxmin}}
\end{figure*}

\begin{figure*}
\includegraphics[width=16.0cm,trim={3mm 0 0mm 0},clip]{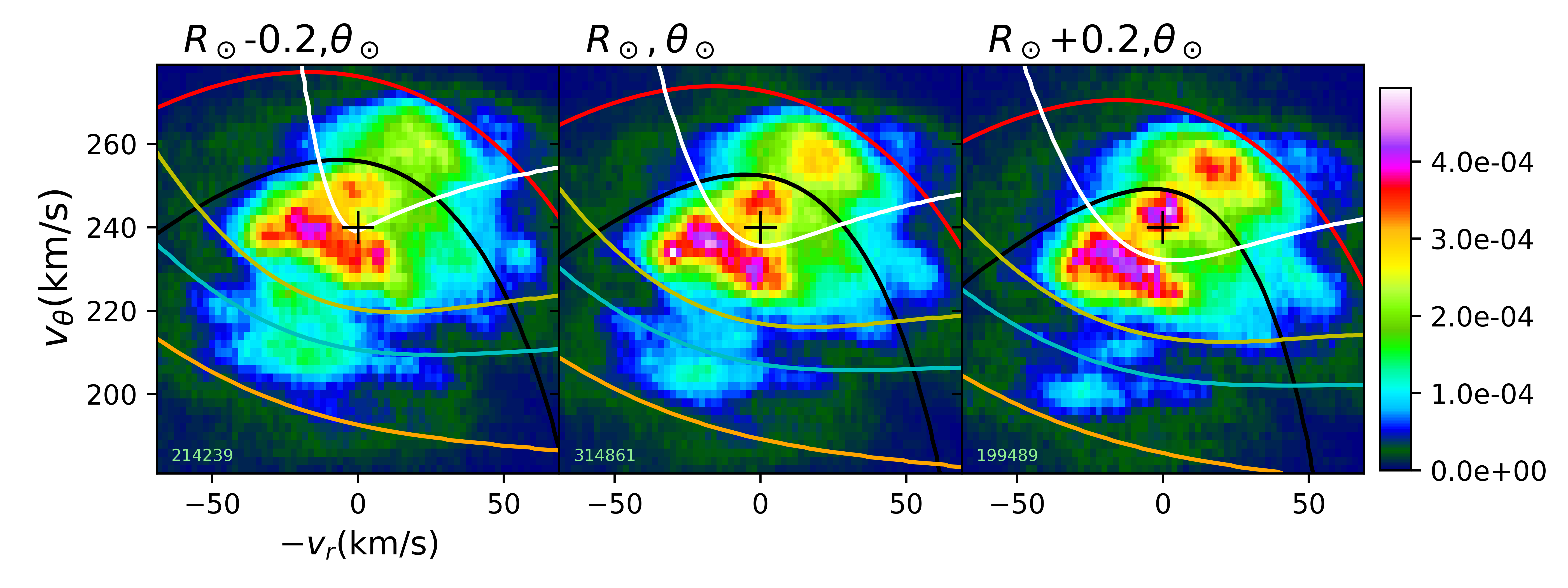}
\includegraphics[width=16.0cm,trim={3mm 0 2mm 0},clip]{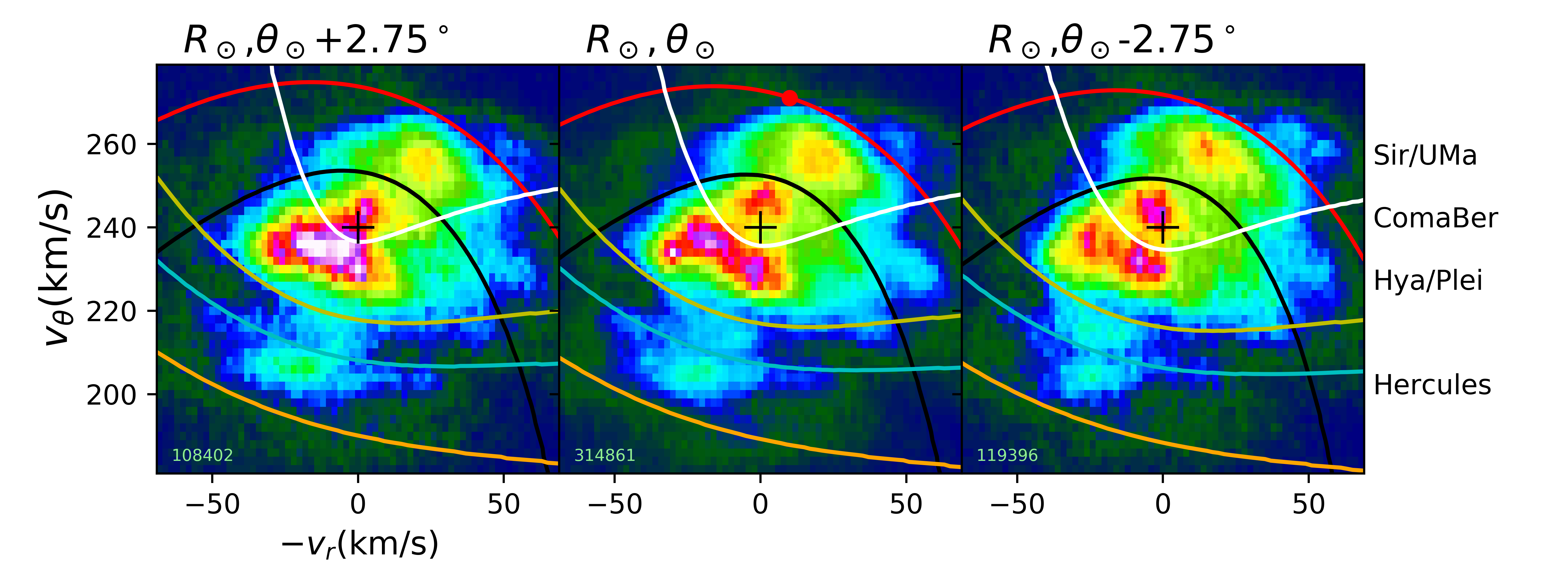}
\caption{Top row:  The color maps show stellar velocity distributions computed from
histograms of {\it Gaia} DR2 data for stars in neighborhoods centred
at $\theta_0$ and at 3 different galactocentric radii.    The coordinate centres of
each neighborhood are listed on the top of each panel.   Using the centres of these neighborhoods
as orbit initial conditions, 
orbits are integrated backwards to their first pericentre or apocentre.  The spiral patterns with properties listed
in Table 1 are touched at pericentre or apocentre on the colored contour lines.  Each contour
corresponds to a particular spiral arm pattern. The colorbar on the right  applies to all six panels.
The total numbers of star in each histogram is written on the lower left in each panel.
A black plus is located at (0,240) \kms in each panel, corresponding to a circular orbit.
Bottom row: Similar to the top row except for neighborhoods centred at $R_\odot$ and at three different
azimuthal angles.    Clumps in the velocity distribution, called moving groups, are labelled on the right.
Orbits that divide the moving groups have apocentres or pericentres that have recently grazed
spiral arms.  The red dot and red line on the lower middle panel shows the orbit and spiral
pattern illustrated in Figure \ref{fig:illustration}.
\label{fig:many}}
\end{figure*}

\section{Crossing spiral arms}
\label{sec:spiral}
   
To estimate the extent of radial or epicyclic
motions for stars in the solar neighborhood, we integrate
orbits backwards, starting at the position of the Sun,
for different current values of radial and tangential velocity components.  
For consistency with the \citet{katz18}, we adopt local standard of rest rotational velocity $v_c = 240$ \kmsns,
and the Sun's galactocentric radius $R_\odot =8.34 $ pc, following \citet{reid14} 
(and see the review by \citealt{bland16}).
Orbits are restricted to the Galactic plane and integrated using a flat rotation curve with rotation
velocity $v_c$ equal to that of the local standard of rest.
The time since the last apocentre and pericentre are shown in color in Figure \ref{fig:ecc} in the 
left and right panels, respectively.   Contours show 
the galactocentric radii of apocentres and pericentres and they are labelled in kpc. 
Figure \ref{fig:ecc} shows that stars at $v_\theta \sim 260 $ \kms could have touched 
the Perseus arm at apocentre.  The  red line labelled  7 kpc on the right panel shows stars that could graze
 the Sagittarius arm at pericentre.  Spiral arms are not fixed. To estimate which 
stars could have crossed them we need to take into account their motion through the Galaxy.

In polar coordinates,
we describe a logarithmic spiral arm density peak  moving through the Galaxy disc 
with pattern speed $\Omega_s$
using a function $R(\theta,t)$
\begin{equation}
\alpha \ln  \left( \frac{R(\theta, t)}{R_{s0}}\right) =  \theta - \theta_\odot -\Omega_s t . \label{eqn:spiral}
\end{equation}
The arm peak has radius $R_{s0}$ at $\theta = \theta_\odot$ at time $t=0$.
We take $t=0$ to be the current time.
The arm pitch angle $p= \arctan \alpha^{-1}$ is the angle between the spiral and the tangent 
to a centered circle of radius $R$ at point $R(\theta)$ on the arm.
We put the Sun at $\theta_\odot = \pi$ and radius $R_\odot$.
The Galaxy rotates clockwise with $\dot \theta <0$. 
A trailing arm has $\alpha >0$ and $\Omega_s <0$.  We do integrations
with negative $v_\theta, \Omega_s$, however, we neglect their signs in our figures and table.

For each tangential velocity $v_\theta$ and radial velocity  $v_r$ at the current time,
we integrate backwards until the orbit reaches either pericentre or apocentre.  
Orbits are restricted to the Galactic plane.
We record the time, azimuthal angle $\theta$ and galatocentric radius of this pericentre or apocentre.
For apocentre these give $t_{apo}, \theta_{apo}$ and $R_{apo}$ consistent with the chosen 
velocity vector.
For winding parameter $\alpha$ and 
pattern speed $\Omega_s$ describing a spiral pattern, we invert equation \ref{eqn:spiral}
using time,  $t=t_{apo}$, azimuthal angle, $\theta=\theta_{apo}$ and radius,  
$R(\theta_{apo},t_{apo})=R_{apo}$,
to solve for $R_{s0}$ consistent with the spiral density peak being at 
the apocentre location at $t_{apo}$.   The radius $R_{s0}$ is  
the radius of the spiral arm peak at the Sun's azimuthal angle
at the current time.  Thus we find the offset of a spiral arm at the current time 
that would have been touched previously
at apocentre  for a star with the chosen velocity vector.    
We repeat the procedure, computing recent orbital pericentres and  radii
$R_{s0}$ for spiral arms that could have been touched at a recent pericentre.

Figure \ref{fig:illustration} shows an illustration of our backwards integration estimate.
A star currently in the solar neighborhood with velocity components
 $v_r=-10$ \kms and $v_\theta = 271$ \kms is shown as a black dot
at the location of the Sun and the star's past orbit is shown with a dotted black line.  
 We integrate its orbit  backwards finding that it
reached apocentre approximately 83 Myr ago at the location of the black dot on the lower right.
A spiral arm like the Perseus arm currently lies outside the radius of the Sun,
however 83 Myr ago, this arm could have grazed the star's orbit when it reached apocentre.
The current and three past locations of the spiral arm are shown as thick cyan lines on the same Figure.
The star only grazes the spiral arm and does not cross it.
A star with a lower initial $v_\theta$ remains inside this  spiral arm during the same time period. 

In  section \ref{sec:comp} we compare curves of constant $R_{s0}$ as a function 
of  $v_r,v_\theta$ together with stellar velocity distributions.   
 The curves in Figure \ref{fig:maxmin} and \ref{fig:many}
are $R_{s0}$ curves, and the colormaps show
local velocity distributions constructed from {\it Gaia} DR2 data.
For the curves in Figure \ref{fig:maxmin} we begin the backwards orbit integrations
 at the location of the Sun (galactocentric radius
$R_\odot$, azimuthal angle $\theta_\odot$ and $z=0$).   
For the curves in Figure \ref{fig:many} we begin the backwards orbit integrations at 4 additional positions;
at $r = R_\odot \pm 0.2 $ kpc, $\theta = \theta_\odot$, 
and at $r = R_\odot$, $\theta =  \theta_\odot \pm 2.75^\circ$.  For these
last two positions, the distance from
the Sun along the direction of rotation is $\pm$ 0.4 kpc.

\subsection{{\it Gaia} DR2 velocity distributions}
\label{sec:data}

The curves in Figures \ref{fig:maxmin} and \ref{fig:many}
are shown with {\it Gaia} DR2 velocity distributions of stars (as color images). 
We use stars  with {\it Gaia} DR2 available radial velocities.
The 3-dimensional velocity components
are  computed using  inferred distances, {\it Gaia} DR2 positions, proper motions and
line of sight radial velocities and with adopted solar peculiar velocity.  
For consistency with the \citet{katz18}, and to compute $v_r, v_\theta, v_z$ from proper motions, distances
and radial velocity,
we adopt the  solar peculiar motion (with respect to the local standard of rest)  by \citet{schonrich10};
$(u_\odot,v_\odot,w_\odot) = (11.1, 12.24, 7.25)$ \kmsns.

The best way to compute distances from parallaxes giving unbiassed results is 
discussed in the current literature (see the review in \citealt{luri18} or \citealt{bailerjones18}).
We compute the stellar distances using a Bayesian scheme which takes the {\it Gaia} DR2 parallax and approximate $G_{\rm RVS}$ magnitudes as input \citep{mcmillan18b}. The estimates adopt a stellar density prior 
(taken from \citealt{burnett10}) and a prior on the absolute $G_{\rm RVS}$ magnitude, derived from the PARSEC isochrones \citep{marigo17}, under reasonable assumptions of the distribution of age, mass and metallicity of stars (also see \citealt{mcmillan18}).
We have also computed stellar distances and extinctions with the {\it StarHorse} code 
\citep{santiago16,queiroz18}, using the parallaxes and $G, G_{\rm BP}, G_{\rm RP}$ magnitudes from {\it Gaia} DR2, together with 2MASS, AllWISE, and APASS photometry, as input. This code calculates the marginal distance posterior probably density function over a grid of PARSEC 1.2S stellar models \citep{Bressan2012,Chen2014,Tang2014}, under reasonable priors for the density, stellar initial mass function, 
and star formation rate of the main stellar components of the Milky Way 
(following \citealt{bland16}; see \citealt{queiroz18} for details), 
and assuming the extinction curve by \citep{Schlafly2016}.
For the nearby stars used in Figures \ref{fig:maxmin} and \ref{fig:many},
the differences in the velocity distributions using distances from {\it StarHorse} compared
to those from McMillan's code 
are insignificant (the differences are barely noticeable
as slight pixel variations). 

The velocity distributions in our figures are computed by making a histogram, counting the
numbers of stars in bins that are 2 \kms square. 
The color-bar on the righthand side of the plots shows numbers of stars per velocity
bin, but normalized so that the histograms integrate to 1.   
The stellar velocity distribution shown in Figures  \ref{fig:maxmin} is comprised of stars
with galactocentric radius $R_\odot - 0.15\ {\rm kpc} < r < R_\odot + 0.15$ kpc
and $\theta_\odot - 1.37^\circ < \theta < \theta_\odot + 1.37^\circ $ (maximum distance along
the direction of rotation from the Sun is 0.2 kpc).  
Star distances from the Galactic plane are restricted to $|z|<0.2$ kpc.
Because we consider only motions in the Galactic plane, we only count stars with $|v_z| < 20$ \kmsns.
The velocity distributions shown in Figures \ref{fig:many} are comprised of stars from
neighborhoods that have the same radial and angular widths as those in Figures \ref{fig:maxmin}.
The centres of the neighborhoods of stars counted in the histograms
are the same as the initial conditions for the orbit integrations and
are labelled on the top of each figure panel.

\subsection{Spiral crossing curves on {\it Gaia} DR2 velocity distributions}
\label{sec:comp}

Figure \ref{fig:maxmin} contours shows spiral radii $R_{s0}$ (at $\theta_\odot$) for possible
spiral arms touched at previous apocentres (top two rows) and 
those that would be touched at pericentres (bottom two rows).    Contours are labelled in kpc.
The top and third row panels show the effect of
varying the spiral pattern speed, and the second and bottom panels show the effect of varying the spiral
pitch angle.
Pattern speeds are given in units of \kmsikpc and listed on top of the panels.
Large or small $R_{s0}$ compared to $R_\odot$ imply that pericentre or apocentre
took place about 100 Myr ago (see Figure \ref{fig:ecc}), not  that the spiral arm
is confined to large or small radii (equation \ref{eqn:spiral} allows the
spiral arm to reach all radii). 
Due to differential rotation in the Galaxy,
a spiral arm with pattern speed that
differs from the angular rotation rate of the Sun ($\Omega_\odot \approx 29$ \kmsikpcns) can more
quickly moves away from the Sun than one with  $\Omega_s \sim \Omega_\odot$.

In Figures \ref{fig:maxmin} a single spiral arm would be expected to cause a change
in the velocity distribution along one of the curves (of constant $R_{s0}$) as stars on one side of
the curve don't cross the arm  and those on the other side do.
Because the location of a spiral arm depends on angle and radius,
the curves on the plots are not symmetric about the $v_r =0$ line.
We find that these curves can exhibit the slopes of arcs and streaks
 in the solar neighborhood's stellar velocity distribution.   

Figure \ref{fig:maxmin} shows that spiral arm crossing curves depend on  arm
pattern speed and winding angle.    We see that variations in pattern speed or winding angle
cause larger differences in radius $R_{s0}$ for orbits that have higher eccentricity.
This is expected because $R_{s0}$ is the present radius of the spiral arm at $\theta_\odot$ 
and higher eccentricity orbits reach apocentre or pericentre distant from the Sun.
Measurements for the pitch angles of spiral arms in  the Galaxy
are $p\approx 12^\circ$ \citep{russeil03,xu16,vallee17}.
 Estimated pitch angles tend to vary or be uncertain by
a few degrees for a particular arm, though there is larger scatter for models that link arms 
or assume a particular number of arms is present (e.g., \citealt{vallee17}).   
In contrast, pattern speed estimates vary widely \citep{naoz07}.
Uncertainties in pattern speed  more strongly
affect our ability to estimate the current radius of a spiral arm $R_{s0}$ (from a
feature in the velocity distribution) than uncertainties in winding angle. 
Hereafter we adopt a winding parameter $\alpha=5$ corresponding to 
pitch angle  $p=11.3^\circ$.  This is approximately consistent with recent pitch angle estimates
for spiral arms near the Sun \citep{xu16}.

So far we have discussed  orbits computed in a logarithmic potential giving  a flat rotation curve.
A rotation curve with rotational velocity $v(r) = v_c \left( {r}/{R_\odot} \right)^\beta$
has slope 
dependent on an exponent $\beta$ (see \citealt{dehnen99}).  With $\beta \to 0$ a flat
rotation curve is recovered.  With $\beta = -0.1$ (approximately consistent
with measurements of Oort's constants; \citealt{bovy17}) the most extreme apocentres
and  pericentres are about 0.5 kpc larger 
than shown on the upper part of Figure \ref{fig:ecc} left panel and the lower part of  Figure \ref{fig:ecc} right panel.
For $\beta =-0.1$, the curves on the in Figure \ref{fig:maxmin} are similar but have radii 
about +0.5 kpc larger in the high eccentricity regions (near the outer boundaries) of the plots.  
 	
\subsection{Multiple Arms}
\label{sec:multiple}

By adjusting the pattern speeds we have identified values of spiral arm positions $R_{s0}$ that could
account for features seen  in the {\it Gaia} velocity distributions. 
Six spiral patterns, specified by their pattern
speed and current radius at $\theta_\odot$ (and with winding parameter $\alpha=5$) are listed in Table 1.
These six patterns are shown as colored contours in Figure \ref{fig:many}.
In Figure \ref{fig:many} velocity distributions and spiral orbit crossings are computed at
5 different regions in the galaxies with central positions at and near the Sun.  
The coordinates of the regions are
written on the top of each panel.  
Orbit integrations used to compute the curves are begun at the central coordinate values
for each region.
The names of common moving groups (or streams) are shown on the right hand side of the  bottom
row of panels  with the height of the label
 matching the $v_\theta$ value of their associated peak in the velocity distribution.  
We have labelled 
the Hercules, Sirius/UMa, Coma Berenices and Hyades/Pleiades streams,
adopting the names used by \citet{dehnen98} (we note that 
their assumed local standard of rest differs from that used here).

Figure \ref{fig:many} displays
stellar velocity distributions at different positions in the Galaxy near the Sun. 
The {\it Gaia} DR2 distributions confirm that the velocities of peaks in 
 the stellar velocity distribution depend on position in the Galaxy \citep{monari18}, and as
shown by previous studies \citep{antoja14,antoja15,monari17b,quillen18}.
In Figure \ref{fig:many}  we see
that peaks in the velocity distribution shift downward by a few \kms with increasing neighborhood 
galactocentric radius.
Even though the distance between the neighborhoods is larger 
in the azimuthal direction than the radial one (0.4 kpc rather than 0.2 kpc), 
variations between the velocity distributions are less pronounced  in the
azimuthal direction (bottom row as compared to the top row of Figure \ref{fig:many}).
The tendency of a gap in the local velocity distribution to shift with radius of the neighborhood
was also seen in the simulations by \citet{quillen11}.

A strong dependence of the morphology of velocity distribution
on galactocentric radius is expected for most dynamical processes.  
With  spiral structure the gap shifts
because stars at different angular momentum values 
can reach different spiral arms.
For phase wrapping, this is expected
because the epicyclic oscillation period is strongly dependent on radius.  
For resonant processes
the strong radial dependence is expected 
because orbital frequencies are dependent on radius. 
In the case of the gap separating the Hercules stream from 
more nearly circular orbits, this was observed with observational data 
\citep{antoja14,monari17b,quillen18}, 
and predicted with models of the bar's resonant effects on the local velocity distribution \citep{dehnen00,monari17a}.

In Figure \ref{fig:many} the red curves correspond to a spiral arm with current radius $R_{s0} = 10 $ kpc
at $\theta_\odot$.
This arm is consistent with  the Perseus spiral arm, about 2 kpc outside the Sun's galactocentric radius.  
In the velocity distributions, we infer that this arm is responsible for the 
outer boundary of the Sirius/UMa moving group, where stars can just reach the arm peak when
they near apocentre.  Figure \ref{fig:illustration} shows the orbit of a star that grazes this arm
at at previous apocentre.
The boundary depends on position in the Galaxy, decreasing in $v_\theta$
at larger galactocentric radius. 
The red line matches the boundary
seen in the local velocity distributions 
for all 5 neighborhoods shown in Figure \ref{fig:many}.  We estimate a pattern speed of 20 \kmsikpc
for this arm with an uncertainty of about $\pm$ 3 \kmsikpcns.  This places the pattern speed below but
near corotation 
at the galactocentric radius of the arm  (at $r=10$ kpc, the angular rotation rate 
$\Omega \approx 24$ \kmsikpcns).

The black contours in Figure \ref{fig:many} correspond to a spiral arm that is nearer the Sun and 
has pattern speed similar to
angular rotation rate of the local standard of rest (that is about 29 \kmsikpcns).  
This arm is consistent with
the Local Spiral arm if it is a separate arm  \citep{xu13,xu16} rather than a spur \citep{houhan14}.
In the velocity distributions, this arm causes a division between the Coma Berenices
and the Sirius/UMa moving groups.  Stars below the black contour curve never reach the
arm, whereas those above it have crossed the arm and recently reached apocentre outside it.

The white curves  correspond to a spiral arm that is currently at the location of the Sun.
This is consistent with a continuation of the Local Spur (see Figure 2 by \citealt{xu16}).
The arm causes a division between the Coma Berenices
moving group and the Hyades/Pleiades moving groups, with stars in the Hyades/Pleiades
moving groups crossing the Local Spur as they neared pericentre.  The Local Spur is probably weak
at the location of the Sun but could have been stronger 60 Myr ago when these pericentres took 
place (times can be estimated using Figure \ref{fig:ecc}).

The yellow and cyan curves correspond to arms with  larger pattern speeds, consistent with spiral
arms closer to the Galactic centre.
The current radii of these arms are both near that expected for the Sagittarius arm.  There is confusion
due to the  presence of an orbital resonance with the Galactic bar 
(e.g., see discussion by \citealt{quillen18} and citations listed in our introduction).  
Our simple orbit integrations do not take into 
account perturbations by the bar.
The lower orange curves, bounding the lower part of the Hercules stream ($v_\theta \sim 205 $ \kmsns),
correspond to a fast spiral pattern in the inner Galaxy with no obvious known
 counterpart.   
We have considered the possibility that a spiral arm might be touched at the second previous pericentre  
(rather than just the last one).  For $\Omega_s = 35$ \kmsikpc and $\alpha=5$
this hypothesis gives curves near the cyan and orange ones that are less tilted  (closer to symmetrical about
the $v_r=0$ line) than the cyan and orange curves and so might better match the orientation 
of the lower part of the Hercules stream.
However, this hypothesis requires predicting orbits more than 200 Myr in the past.
Until we have a good model for the recent behavior of spiral arms and have pinned down the bar
pattern speed, we probably
can't  predict the locations and times of pericentres well enough to relate it to spiral structure that might have
been present that long ago. 
 
Figure \ref{fig:many} suggests that nearby spiral arms have recently
influenced the local velocity distribution. 
The Lin-Shu hypothesis \citep{linshu} postulates that
spiral arms are long lived patterns.
However, N-body simulations can 
show spiral arm features which are short-lived but recurrent (transient) 
and with pattern speeds that  match the rotation of the stars, 
(approximately co-rotating; e.g., \citealt{sellwood84,grand12,roca13,kawata14}).
Or multiple patterns might exist simultaneously with interference between them 
giving transient-like behavior 
(e.g., \citealt{quillen11,comparetta12,minchev12}).
The match between velocity distribution boundaries with recent spiral structures 
would not be expected if spiral arms are long lived, as 
the  distribution function would be  relaxed and more like that computed by \citet{monari16}.
A nearly corotating spiral structure causes stronger perturbations on
stars reaching it near apocentre than one that has a slower pattern speed \citep{kawata14,hunt17}.
Our estimated pattern speeds (in Table 1) tend to be faster in the inner Galaxy than
the outer Galaxy so are consistent with nearly corotating transient structures.
Simulations have illustrated that transient spiral structures can migrate stars
(e.g., \citealt{grand12,comparetta12}).  Large additional numbers of higher eccentricity tracer particles would be
required in a simulation to determine if there are associated arcs in local velocity distributions.

\begin{table}
\vbox to 95mm{\vfil
\caption{\large  Spiral arm boundaries
\label{tab:arms}}
\begin{tabular}{@{}llllll}
\hline
Speed  &  radius   & peri/apo  &  color &  arm\\
$\Omega_s $   &   $R_{s0}$   &                &           &  \\
\hline
20 & 10.0   & apo & red & Perseus \\
27 &  9.2 & apo &  black & Local Arm \\
29 & 8.0 &  peri &  white & Local Spur  \\
33 & 7.0 & peri & yellow  & Sag, BR \\
35 & 6.5 &  peri & cyan &  Sagittarius \\
45 & 6.1 & peri & orange &  ? \\
\hline
 \end{tabular}
{\\  Notes.  Each colored curve (from top to bottom) 
shown in Figure \ref{fig:many} corresponds to a spiral arm
listed here in corresponding order.   The colors of the curves in the figures are listed in column 4.
Each spiral arm is described with equation \ref{eqn:spiral} and depends on a pattern
speed, $\Omega_s$ in \kmsikpcns, a winding parameter,  
$\alpha = 5$, corresponding to a pitch angle of $11.3^\circ$,
and the present radius of the arm peak $R_{s0}$ in kpc at $\theta_\odot$.
We also specify whether the curve in Figure \ref{fig:many} is for pericentre or apocentre crossing.
The rightmost column lists associated known arms in the Galaxy near the Sun, referring
to the map in Figure 2 by \citet{xu16}.
BR stands for a resonance with the Galactic bar.  
Sag stands for the Sagittarius arm.
The boundaries are estimated using orbits computed in a potential with a flat rotation curve.
}}
\end{table}

\section{Summary and Discussion}
\label{sec:sum}

Disc stars near the Sun have crossed spiral arms.
We propose that  divisions or boundaries  in local velocity 
distributions separate stars that have recently crossed or been strongly perturbed by 
a particular arm from those that haven't. 
This scenario is similar to that explored  by \citet{hunt17} for high angular momentum disc stars
in the solar neighborhood
with apocentres approaching the Perseus spiral arm.
By estimating the time, angle and radius of recent pericentres and apocentres and using
a logarithmic spiral model,
we estimate pattern speeds and radii of spiral patterns that could have formed
boundaries in stellar velocity distributions constructed from nearby stars.
The model matches  tilts of some arcs present in the $v_r,v_\theta$
local velocity distributions constructed from {\it Gaia} DR2 data and how these arcs 
vary with position in the Galaxy. 
The model accounts for the outer edge of the
Sirius/UMa moving group with recent apocentres nearing the Perseus spiral arm.   
The division between
the Coma Berenices and Sirius/UMa moving groups traces apocentres near the Local Spiral arm, 
and the division between Coma Berenices and Hyades/Pleiades moving groups with
pericentres near the Local Spur, with Local arm and spur locations as mapped by \citet{xu16}.   
This model requires multiple spiral patterns present nearby in the Galaxy, supporting
a more flocculent rather than a grand design morphology for the outer Galaxy (e.g., \citealt{quillen02,xu16}).

Our model arm crossings do not accurately predict the morphology of streaks in the Hercules stream
and their dependence on position in the Galaxy.
The scenario is less successful for faster patterns present in the inner Galaxy, possibly because
we have not taken into account perturbations from the Galactic bar. 
We failed to find curves that bound
the clump in the velocity distribution at $v_\theta \sim 230$ and $-v_r  \sim 60 $ \kms 
known as the Wolf 630 moving group.  
Solar neighborhood
stars (such as in the Hercules or Wolf 630 moving groups)  
could have been perturbed by spiral arms that are now distant from the Sun, making it harder to match
structures in the velocity distribution at higher orbital eccentricity with known spiral structures.

We have proposed a relation between boundaries in local velocity distributions and loci of
pericentres or apocentres that have recently approached spiral arms.  
Without doubt spiral arms perturbations influence local velocity distributions, but how they do so
is likely more complex than defined by a simple pericentre or apocentre locus.
The velocity perturbation on a star caused by grazing an arm is likely to be similar to 
the size of bumps and wiggles ($\pm$ 10--20 \kms \citealt{mcgaugh16}),
measured in the Milky Way rotation curve caused by  spiral arms.
The velocity perturbation would also be sensitive to the strength, width and pattern speed of
the arm and the length of time that the star is near the spiral's density peak.  
Orbit integrations that take into account the perturbation of the arm are needed
to quantitatively estimate the depth of a feature in the velocity
distribution caused by close passages to a spiral arm.
Simulations containing many particles  
are needed to match the numbers of stars in the {\it Gaia} database 
and resolve structure in phase space at higher eccentricity while simultaneously 
allowing self-consistent self gravitating spiral structures  to develop in the colder disc population. 
Time dependent and non-equilibrium models with spiral arms could be improved  for 
the stellar phase space distribution functions.    Structure and gradients in 
the velocity distribution imply that there are variations with position 
in the number density of stars.  With improvements 
in 3D extinction maps and better statistical models, we may improve maps of  
spiral structure in the Galaxy. 
These developments would be needed to
confirm or explore the connection between boundaries in local velocity distributions and spiral arms.

Spiral structure alone cannot account for the vertical motions studied by \citet{antoja18}
or vertical gradients in the velocity distribution, as
illustrated by the Coma Berenices moving group \citep{quillen18,monari18}.
High angular momentum and high eccentricity stars and those with large vertical amplitudes 
should be less strongly affected 
by spiral arms and might more prominently 
show phase wrapping associated with perturbations in the outer Galaxy, 
and affecting both planar and vertical velocity distributions. 
Warp or spiral-like features can be induced by external perturbations
\citep{delavega15, chequers17} and these might have more strongly perturbed stars
with particular vertical and horizontal epicycles, causing substructure
in velocity distributions as a function of $v_z$ as well as $v_r$ and $v_\theta$.
In our model, boundaries in velocity distributions 
are not solely dependent on particular orbital periods, differing 
from a resonant model where a particular period sets a 
gap in phase space (e.g., see \citealt{quillen18}).   
The influence of spiral structure on local velocity distributions  may blur gaps caused by  
orbital resonances in the action distributions (see those computed by \citealt{trick18}).  

\vskip 2 truein

A.C. Quillen is grateful to the Leibniz Institut f\"ur Astrophysik Potsdam (AIP) for their
warm welcome, support and hospitality July 2017 and April-May 2018.
A.C. Quillen thanks Mt. Stromlo Observatory
for their warm welcome and hospitality Nov 2017-- Feb 2018.
A. C. Quillen is grateful for generous support from the Simons Foundation and her work
 is in part supported by NASA grant 80NSSC17K0771.
I. Minchev acknowledges support by the Deutsche Forschungsgemeinschaft under the grant MI 2009/1-1.

We thank Anna Queiroz and Basilio Santiago for help with {\it StarHorse}.
We thank Gal Matijevic  for invaluable and continued support.
We thank the E-Science and Supercomputing Group at Leibniz Institute  
for Astrophysics Potsdam (AIP) for their support with running the {\t StarHorse} code on AIP cluster resources.

This work has made use of data from the European Space Agency (ESA) mission
{\it Gaia} (\url{https://www.cosmos.esa.int/gaia}), processed by the {\it Gaia}
Data Processing and Analysis Consortium (DPAC,
\url{https://www.cosmos.esa.int/web/gaia/dpac/consortium}).  Funding for the DPAC
has been provided by national institutions, in particular the institutions
participating in the {\it Gaia} Multilateral Agreement.


\end{document}